\documentstyle[titlepage,11pt]{article}
\begin{document}
\begin{center}
\bigskip
\bigskip
{\bf Piezoelectricity of Cholesteric Elastomers}\\
\bigskip
\bigskip
\bigskip
Robert A. Pelcovits$^\ast$\\
\medskip
The Martin Fisher School of Physics \\
Brandeis University \\
Waltham,
Massachusetts 02254\\
and\\Department of Physics \\
Brown University \\
Providence, RI 02912$^\dagger$\\
\medskip
\bigskip
\medskip
Robert B. Meyer\\
\medskip
The Martin Fisher School of Physics \\
Brandeis University \\
Waltham,
Massachusetts 02254
\end{center}
\vfill
\begin{flushleft}
Submitted to Journal de Physique II, Chemical Physics
\medskip
$^\ast$ Author to whom correspondence should be addressed at:
bob@het.brown.edu\\
\dag Permanent address
\end{flushleft}
\break

\begin{abstract}

We consider theoretically the properties of piezoelectricity in cholesteric
elastomers. We deduce using symmetry considerations the piezoelectric
contributions to the free energy in the context of a coarse-grained description
of the material. In contrast to previous work we find that compressions or
elongations of the material along the pitch axis do not produce a piezoelectric
response, in agreement with fundamental symmetry considerations. Rather only
suitable shear strains or local rotations produce  a polarization. We propose
some molecular mechanisms to explain these effects.

\end{abstract}

Piezoelectric materials are characterized by the appearance of an
electric polarization when a suitable mechanical stress is applied\ [1].
Symmetry considerations require that these materials be non-centrosymmetric,
i.e. not invariant under inversion. Chiral liquid crystals such as a
cholesteric or smectic $C^*$ satisfy this requirement but their fluid nature
will not support a static shear. However, chiral liquid crystalline elastomers
which consist of a cholesteric liquid crystal homogeneously embedded in a
polymer gel can support static stresses, including shear, due to the presence
of the
underlying gel structure. Thus, they are candidates for the observation of true
piezoelectricity in a liquid crystalline system.

In this paper we consider theoretically the nature of piezoelectricity in a
cholesteric elastomer. We show that a shearing of the elastomer along
 the pitch axis causes  a piezoelectric response. Local rotations of the
elastic medium can also in principle produce a polarization.
In contrast to previous work\ [2,  3], we find that compressions or elongations
of the elastomer along the pitch axis cannot produce a polarization. We also
propose some molecular arguments  to explain the mechanism of piezoelectricity
in these materials.Finally we  discuss the relationship of our work to previous
 experimental studies\ [4,  5].

Our starting point is a hydrodynamic description of cholesterics in terms of a
pitch vector due to Lubensky\ [6]. While it is possible to develop a
description
 in terms of the director, all macroscopic quantities such as the
polarization require a coarse-graining, i.e. an averaging of the director field
over the pitch length.
The description in terms of the pitch vector is already coarse-grained. We
demonstrate below that our theory is equivalent to Terentjev's theory of
piezoelectricity\ [3] which is phrased in terms of the director. However, the
coarse-graining built into the pitch vector description is more convenient in
determining the elastic strains associated with piezoelectricity.  The director
$\bf n$ in a cholesteric has the following form\ [6]:

$$
\bf n (\bf r) =\bf n_{\rm 0} \cos \psi (\bf r) + \bf p \times \bf n_{\rm 0}
\sin \psi (\bf r),
\eqno(1) $$
where $\bf p$ is a unit vector along the pitch axis, $\bf n_{\rm 0}$ is a unit
vector in the plane perpendicular to $\bf p$, and $\psi$ is the phase
angle of the director. The latter quantity may be expressed as $\psi =
{2\pi\over\lambda}\bf p \cdot \bf r + \phi$, where $\lambda$ is the pitch of
the helix, and $\phi$ is a phase factor. We can define a wavevector $q_0$ for
the helix via the relation, $q_0={2\pi \over \lambda}$. It is important to note
that a helix is not a polar object, i.e. it looks the same whether viewed from
the top or the bottom. Mathematically speaking this nonpolarity arises from the
existence of twofold axes parallel to $\bf n$ and $\bf p \times \bf n$.
Because the helix is a nonpolar object, then as Lubensky first noted, $\bf p$
can be chosen to be either a vector or a pseudovector. The direction of $\bf p$
 along the helical axis is not a physically relevant quantity. If $\bf p$ is
chosen to be a vector (pseudovector) then
the phase angle  $\psi$ is  a pseudoscalar (scalar). In either
case $q_0$ is a pseudoscalar. These symmetry considerations must be borne in
mind when constructing free energies or hydrodynamic equations, and will ensure
that the helix is treated as a nonpolar object. Providing these symmetries are
respected, the coarse-grained theory is completely equivalent to the director
theory, contrary to recent assertions\ [3,  7]. The elastic degrees of freedom
of the gel are described by a symmetric strain tensor $u_{ij}={1 \over 2}
\bigl({\partial u_i \over \partial x_j} +  {\partial u_j \over \partial
x_i}\bigr) $ as well as a
rotation
pseudovector ${\bf \omega} = {1 \over 2}{\bf \nabla} \times \bf u$, where $\bf
u$ is
the displacement field of the network.

We now use these symmetry considerations to construct the form of the
piezoelectric contribution $F_p$ to the free energy density,  i.e. the
contribution linear in the applied electric field $\bf E$ and the strain. We
find the following expression:

$$
F_p = \gamma_1 q_0 E_i \epsilon_{ijk} p_j p_l \delta_{km}^{(tr)}u_{ml} +
       \gamma_2 q_0 E_i p_i p_j \omega_j +
       \gamma_3 q_0 E_i \delta_{ij}^{(tr)} \omega_j
\eqno(2) $$
where $\gamma_{1,2,3}$ are the piezoelectric coefficients, $\epsilon_{ijk}$ is
the antisymmetric pseudotensor, and $\delta_{ij}^{(tr)}$ is the projection
operator
transverse to the pitch axis $\bf p$, defined by:  $\delta_{ij}^{(tr)} \equiv
\delta_{ij} - p_i p_j$. Repeated indices are to be summed over. Each of these
terms is a scalar irrespective of whether we choose $\bf p$ to be a vector or
pseudovector. The three terms appearing on the right hand side of equation (2)
are
the only scalars that can be constructed linear in $\bf E$ and the strain
tensor or rotation pseudovector.

The meaning of these terms is as follows. Summing over the indices we see that
the term proportional to $\gamma_1$ involves a shear in the $\bf p$ direction,
e.g. sliding the planes of the elastomer perpendicular to $\bf p$ over each
other. If $\bf p$ is parallel to $\bf \hat z$, then a nonzero $u_{xz}$ strain
component will produce a polarization along the $y$ axis.  There are no
contributions to equation (2) from compressions
or elongations. This result is not surprising because these latter strains will
{\it not} destroy the two-fold rotation axes perpendicular to $\bf p$. The
terms
proportional to $\gamma_2$ and $\gamma_3$ involve local rotations in planes
parallel and perpendicular to $\bf p$ respectively. We provide an intuitive
explanation below for why these rotations can produce a polarization.

Before discussing possible molecular mechanisms for these terms we consider the
relationship of our work to previous studies. Brand\ [2] was the first to study
the piezoelectric response of cholesteric elastomers theoretically. He wrote
down the following
piezoelectric free energy density $F_p^B$:

$$
F_p^B = \zeta_{ijk} q_0 E_i u_{jk}
\eqno(3) $$
where the piezoelectric tensor $\zeta_{ijk}$ has the form

$$
\zeta_{ijk} = \zeta_1 p_i p_j p_k + \zeta_2 p_i \delta_{jk}^{(tr)} +
               \zeta_3 (p_j \delta_{ik}^{(tr)} + p_k \delta_{ij}^{(tr)}).
\eqno(4)$$

Unfortunately, this free energy density is not a scalar because $\bf p$
appears raised to an odd power. This incorrect form led Brand to the physically
erroneous conclusion that an elongation or compression along the pitch axis can
induce piezoelectricity.
Likewise the cholesteric piezoeletric term $p_i p_j E_i \nabla_j \psi$
introduced by Brand and Pleiner\ [8]
and reproduced in ref.~2,   is
also not a scalar. While $\bf p$ appears quadratically, the presence of $\psi$
to the
first power leaves the parity of this term ambiguous. It will be a scalar if we
choose  $\psi$ to be  a scalar, but it will be a pseudoscalar if we  choose
$\psi$ to be a
pseudoscalar. In ref. 2 Brand also considered rotations of the elastic network
relative to the director. His term $(\psi - p_i \omega _i) q_0 p_j E_j$ is a
generalization of our $\gamma_2$ term above to the case where the director can
rotate uniformly and is not hindered by the elastic network. However, his
additional term
coupling this relative rotation to the mass density fluctuatations is again
incorrect on symmetry grounds.

More recently Terentjev\ [3] constructed  free energies for both chiral and
nonchiral  nematic elastomers solely in terms of the director. In the former
case he wrote down the following piezoelectric free energy density:

$$
F_p^T = Q_1 \epsilon_{ijk} E_j n_k u_{il} n_l +Q_2 E_i \omega_i +
       Q_3 E_i n_i \omega_j n_j
\eqno(5) $$

Terentjev noted that whereas Brand has three terms coupled to the symmetric
strain tensor (equation (4)  above), he has only one, the term proportional to
$Q_1$. Terentjev attributed this
discrepancy to Brand's  use of the coarse-grain approximation for cholesterics,
i.e. his use of the pitch axis $\bf p$ rather than the director $\bf n$.
Terentjev claimed that a coarse-grain description is often misleading. However,
we now demonstrate that when equation (5) is coarse-grained {\it properly},
Terentjev's expression becomes identical to our expression, equation (2). Thus,
it
was Brand's incorrect implementation  of coarse-graining discussed above that
was at
fault. Furthermore,
as we discuss below, our correct  coarse-grain theory has some advantage over
Terentjev's approach when applied to cholesterics.  Terentjev coarse-grained
equation (5) incorrectly and concluded erroneously that compressions or
elongations along the pitch axis can produce piezoelectricity.

The proper coarse-graining of Terentjev's expression is easily done if we note
that
using equation (1), the spatial average of the product $n_i n_j$ over a pitch
length
equals $\delta_{ij}^{(tr)}$. Denoting the result of coarse-graining Terentjev's
expression by $\langle F_p^T \rangle$, we find:

$$
\langle F_p^T \rangle = Q_1 E_i \epsilon_{ijk} \delta_{jl}^{(tr)}u_{kl}
                 + Q_2  E_i \omega_i + Q_3 E_i \delta_{ij}^{(tr)} \omega_j
\eqno(6) $$
The term proportional to $Q_1$ is identical in form to our $\gamma_1$ term if
we use the identity:

$$
u_{kl} = \delta_{km} u_{ml} = (p_k p_m + \delta_{km}^{(tr)}) u_{ml}
\eqno(7) $$
When this identity is inserted into the $Q_1$ term, the last term in equation
(7)
does not contribute to the final sum over indices.
Using a similar identity in the term proportional to $Q_2$ we find that
equation (6)
is
identical to our result, equation (2),  with the identifications: $\gamma_2 q_0
=
Q_2$ and $\gamma_3 q_0 = Q_2 +Q_3$.

Terentjev applied his free energy density, equation (5), to a {\it uniform}
director pattern, e.g. a
thin layer (compared to the pitch length)
of cholesteric elastomer perpendicular to $\bf p$. His $Q_1$ term then leads
correctly
to a polarization along $\bf p$ when the layer is compressed or elongated
along that direction. However, he then concluded erroneously that a macroscopic
polarization will arise when one looks at the full pitch. In fact the local
polarization averages to zero over the pitch length, as we have seen from our
$\gamma_1$ term which only involves shear strains. Once the coarse-grained
theory is properly constructed as we have done above in equation (2), it is
readily
apparent that only shear strains are involved, and no errors will creep into
the analysis of piezoelectricity for cholesteric systems.

We now offer some physical arguments to explain the origin of the piezoelectric
terms in equation(2).
The term proportional to $\gamma_1$ involves a shear in the $\bf p$ direction,
e.g. a nonzero $u_{xz}$ strain component, if $\bf p$ is parallel to $\bf \hat
z$.
This shear strain distorts the elastic network. For instance, if the network
was initially isotropic it becomes anisotropic with elongation parallel to an
axis oriented at $45^\circ$ to $\bf p$ in the $xz$ plane. Likewise, even if
the network is intially anisotropic (locally symmetric about $\bf \hat n$) its
axis
of symmetry  rotates towards $\bf p$ through an angle proportional to
$u_{xz}$. In either case the director everywhere will gain a $z$-component by
rotating about the $y$ axis through an angle $\alpha u_{xz}$, where $\alpha$ is
a measure of the coupling of the director to the strain. The value of $\alpha$
will depend on the material structure. For positive $\alpha$ the director will
rotate towards the elongation direction, while for negative $\alpha$ it will
rotate away.   Equation (1) will be replaced by,

$${\bf n} ({\bf r}) = {\bf \hat x} \cos \psi ({\bf r}) + {\bf \hat y}
\sin \psi ({\bf r}) + {\bf \hat z} \cos \psi ({\bf r})\alpha  u_{xz},
\eqno(8) $$
for small values of $u_{xz}$. This director pattern now exhibits splay and bend
as well as the cholesteric twist (see Figure 1). The splay-bend pattern
produces a polarization $\bf P$ due to flexoelectricity\ [9]:

$${\bf P} = e_s {\bf n} ({\bf \nabla} \cdot {\bf n}) - e_b ({\bf n} \times
({\bf
    \nabla} \times {\bf n}))
\eqno(9)$$
where $e_s$ and $e_b$ are the flexoelectric coefficients. Inserting equation
(8)
into equation (9) and averaging over a pitch length we find,

$$\langle {\bf P} \rangle = {\bf \hat y} q_0 \bar e \alpha u_{xz}
\eqno(10) $$
where $\bar e$ is the arithmetic average of $e_s$ and $e_b$. We thus identify
$\gamma_1$ with $\alpha\bar e$.

The terms proportional to $\gamma_2$ and $\gamma_3$ cannot be understood on the
basis of flexoelectricity and require a more novel mechanism. One scenario is
based on the symmetry of the cholesteric molecules, combined with the two
component nature of the cholesteric elastomer. Because of its chiral symmetry,
each cholesteric molecule can be thought of as a miniature propeller blade. A
rotation of the medium surrounding a propeller  produces a unique displacement
of the propeller, just as a rotation of the propeller would produce a
displacement of the medium. This relative displacement of two dissimilar
components produces a polarization. In the most extreme case, the gel and the
cholesteric molecules carry opposite electrical charges, and the polarization
arises from the spatial separation of these charges.

Although the local rotation strains of the gel are valid sources of
piezoelectricity, it is difficult to produce them by simple uniform distortion
of a macroscopic sample. Thus the $\gamma_1$ term will be the most easily
observed piezoelectric effect. Another way to produce relative rotations of the
gel and cholesteric molecules is to apply a torque to the cholesteric
molecules, for instance by using an externally applied magnetic field $\bf H$,
oriented at an oblique angle to $\bf p$. The director will rotate typically
towards $\bf H$, and polarization will be induced in the $\bf p \times \bf
H$ direction, by a combination of both the flexoelectric and the local rotation
mechanisms. Note that the gel maintains the average direction of $\bf p$
during this process; it is only the director which rotates.

Finally, we discuss experiments to observe the piezoelectric effect in these
materials.
Inspired by Brand's theory\ [2] of piezoelectricity in cholesteric elastomers,
two groups\ [4,  5]
undertook expriments.  Following Brand's predictions, they looked for a voltage
to appear
between two plates that compressed a cholesteric elastomer sample in a
direction parallel
to the helix axis.  Had this experiment been performed precisely as stated, the
symmetry
of the cholesteric would have required a null result.  However, both groups saw
measurable
effects, and Meier and Finkelmann\ [5] reported extensive experiments to
characterize these
effects, in relation to variation of the temperature and of the sign and
magnitude of $q_0$.

What effect was seen in these experiments?  In response to a suggestion by one
of us (RBM)
that the sample might not be perfectly symmetric, Meier and Finkelmann reported
that by
altering the shape of the meniscus at one end of the cylindrical samples being
studied,
from negative to positive curvature,
they could change the magnitude and even the sign of
the measured effect.  This suggests that what they were measuring was in fact
the
shear-induced piezoelectric effect described above, arising from the
nonuniformity of
deformation that resulted from compressing the curved end of the sample.  The
precise
geometry of this deformation is unknown, so the resulting voltage produced is
impossible
to predict.

Although a quantitative value of the piezoelectric coefficient cannot be
extracted from
the reported experiments, some of the qualitative characterization of the
piezoelectric
effect in Meier and Finkelmann's work is quite clear.  They did demonstrate
that the
induced voltage is linear in the strain, after an initial deformation that
served to align
the sample.  They also found that the effect is linear in $q_0$, by changing
both the
handedness and the magnitude of the helix pitch.  Both these results agree with
our
theory.

In a properly designed experiment, a monodomain sample of linear dimensions
$dx$, $dy$, and $dz$, with helix axis
in the $z$ direction, would be contained between two parallel plates, located
at $z=0$ and $z=dz$.  A shear strain $u_{xz}$ would be created by displacing
one of the plates relative to the other in the $x$ direction.  This would
induce a piezoelectric polarization in the $y$ direction, which would be
detected by suitably placed electrodes, for instance on the faces of the sample
at $y=0 $ and $y=dy$.  This would provide a direct measurement of $\gamma_1$.
For the inverse effect, a voltage would be applied to the
electrodes, to
induce a shear deformation that would be detected by a relative displacement of
the parallel plates.

\pagebreak

\centerline {\bf Acknowledgements}
\bigskip

R.A.P. wishes to thank the Brandeis University Physics Department for  their
hospitality. R.A.P. was supported in part by the NSF under grant no.
DMR92-17290, and R.B.M. was supported by the ARO through grant no.
DAAL03-92-G-0387 and by the Martin Fisher School of Physics.

\bigskip
\bigskip
\centerline {\bf References}
\bigskip

\begin{enumerate}
\item  See, e.g., L. D. Landau and E. M. Lifshitz,  Electrodynamics of
Continuous Media (Pergamon, Oxford, 1960) pp. 73-79.

\item  H. R. Brand, {\it Makromol. Chem. Rapid Commun.} {\bf 10} (1989)  441.

\item  E. M. Terentjev, {\it Europhys. Letts.} {\bf23} (1993)  27.

\item  S. U. Vallerien, F. Kremer, E. W. Fischer, H. Kapitza, R. Zentel and H.
Poths, {\it  Makromol. Chem. Rapid Commun.}  {\bf11} (1990)  593.

\item  W. Meier and H. Finkelmann, {\it Makromol. Chem. Rapid Commun.}  {\bf11}
        (1990)  599; {\it Macromolecules} {\bf26}  (1993)  1811.

\item  T. C. Lubensky, {\it Phys. Rev. A} {\bf6} (1972) 452.

\item  E. M. Terentjev and M. Warner, {\it J.  Phys. II France} {\bf4} (1994)
849.

\item  H. R. Brand and H. Pleiner, {\it J.  Phys. France} {\bf45} (1984) 563.

\item  R. B. Meyer, {\it Phys. Rev. Lett.}  {\bf22} (1969) 918. In the context
of
cholesterics, see also, J. S. Patel and R. B. Meyer, {\it Phys. Rev. Lett.}
{\bf 58} (1987)
1538.
\end{enumerate}
\pagebreak
\centerline {\bf Figure Caption}
\bigskip

Fig. 1:  The director pattern in a plane cut through the cholesteric helix at
an angle to the helix axis. This plane is rotated about the $y$ axis, through
an angle $\alpha u_{xz}$ relative to the $z=0$ plane.  The director is
parallel to this plane, and exhibits a uniform rotation pattern consisting
of alternating bands of splay and bend.  This produces a flexoelectric
polarization in the $y$ direction.
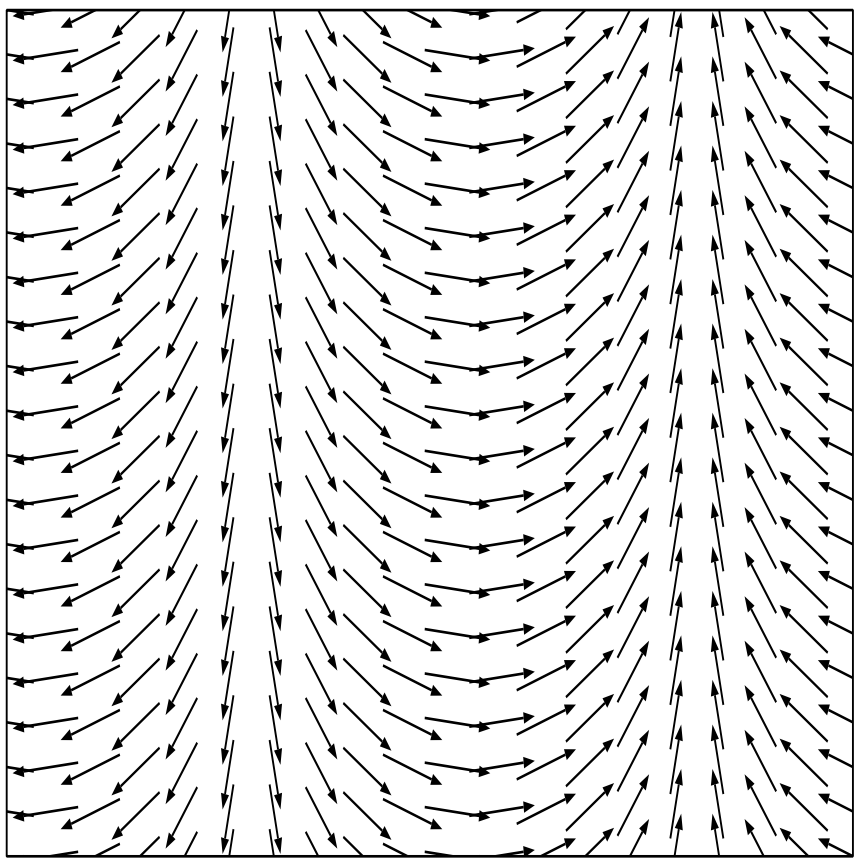
\end{document}